\documentclass[preprint]{aastex}

\input{epsf}

\newcommand{\epsfigx}[1]{\epsfxsize=14cm \epsfbox{#1}}

\slugcomment{Submitted to the ApJ Supplements}

\shorttitle{$BVRI$ photometry of Cepheids in NGC~1866}
\shortauthors{Gieren et al.}

\begin{document}


\title{Cepheid variables in the LMC cluster NGC~1866. I. New BVRI 
CCD photometry}


\author{Wolfgang P. Gieren\altaffilmark{1}}
\affil{Universidad de Concepci\'on, Departamento de F\'{\i}sica,
       Grupo de Astronom\'{\i}a, Casilla 160-C, Concepci\'on, Chile}
\email{wgieren@coma.cfm.udec.cl}

\author{Mat\'{\i}as G\'omez\altaffilmark{1}}
\affil{P.  Universidad Cath\'olica de Chile, Departamento de 
       Astronom\'{\i}a y Astrof\'{\i}sica, Casilla 104, Santiago 22, Chile}
\email{mgomez@astro.puc.cl}

\author{Jesper Storm}
\affil{Astrophysikalisches Institut Potsdam, An der Sternwarte 16,
       D-14482 Potsdam, Germany} 
\email{jstorm@aip.de}

\author{Thomas J. Moffett\altaffilmark{2}}
\affil{Purdue University, Department of Physics, 1396 Physics Building,
       West Lafayette, IN 47907-1396}
\email{moffett@physics.purdue.edu}

\author{L.  Infante}
\affil{P. Universidad Cath\'olica de Chile, Departamento de Astronom\'{\i}a
       y Astrof\'{\i}sica, Casilla 104, Santiago 22, Chile}
\email{linfante@astro.puc.cl}

\author{Thomas G. Barnes III\altaffilmark{2}}
\affil{McDonald Observatory, The University of Texas at Austin, 
       Austin, TX 78712-1083}
\email{tgb@astro.as.utexas.edu}

\author{Doug Geisler}
\affil{Universidad de Concepcion, Departamento de F\'{\i}sisca, 
       Grupo de Astronom\'{\i}a, Casilla 160-C, Concepci\'on, Chile}
\email{doug@kukita.cfm.udec.cl}

\and

\author{Pascal Fouqu\'e}
\affil{Observatoire de Paris, Section de Meudon, 
       DESPA F-92195 Meudon Cedex, France, and European Southern Observatory,
       Casilla 19001, Santiago 19, Chile}
\email{pfouque@eso.org}



\altaffiltext{1}{Guest Investigator, Las Campanas Observatory, 
     which is operated by the Carnegie Institution of Washington}
\altaffiltext{2}{Visiting Astronomer, Cerro Tololo Inter-American Observatory,
                 Chile. CTIO is operated by AURA, Inc., under contract to
                 the National Science Foundation}


\begin{abstract}

     We report $BV(RI)_C$ CCD photometric data for a group of seven
Cepheid variables in the young, rich cluster NGC~1866 in the Large
Magellanic Cloud. The photometry was obtained as part of a program to
determine accurate distances to these Cepheids by means of the infrared
surface brightness technique, and to improve the LMC Cepheid database for
constructing Cepheid PL and PLC relations. Using the new data together
with data from the literature, we have determined improved periods for
all variables. For five fundamental mode pulsators, the light curves are
now of excellent quality and will lead to accurate distance and radius
determinations once complete infrared light curves and radial velocity
curves for these variables become available.

\end{abstract}




\clearpage
\newpage

\section{Introduction and motivation for our study}

     Cepheid variables are the most important local distance calibrators
to lay the foundations of the extragalactic distance ladder. Cepheids in
the Magellanic Clouds, and in particular the Large Magellanic Cloud,
have played a crucial role in this effort ever since
Henrietta Leavitt \citep{Pickering}
discovered the Cepheid PL relation in the SMC in the early years
of the 20th century. Since LMC Cepheids all lie at virtually the same
distance, as opposed to Galactic Cepheids which are found over a range
of distances, and since the LMC is rich in Cepheids and contains Cepheid
variables up to the largest
pulsation periods ($\approx 100$~days), the slopes of the PL and PLC
relations are best established in this satellite galaxy. Furthermore,
the geometrical structure of the LMC is basically a simple tilted
disk with the NE side closest to us \citep{CC86,welch87}, which makes
it possible to apply relatively accurate position-dependent distance
corrections to LMC objects to refer distances to the LMC barycenter,
and thus further reduce the dispersions in the intrinsic Cepheid PL
and PLC relations.  Unlike the slope, the zero point of the PL
relation is much harder to calibrate. This step involves the need to
determine distances to Cepheid variables with methods independent of
the PL relation. The classic approach to solve this problem has been
the use of Cepheids in Galactic clusters whose distances can be derived
from ZAMS-fitting to their observed magnitude-color diagrams. However,
recent HIPPARCOS parallax data on a number of open clusters have
shown that the location of cluster main sequences seems to depend,
in a stronger way than predicted by stellar evolution models, on the
cluster age (e.g. \citet{vanLeeuwen99}), and as long as this dependence
(and perhaps others) is not well understood and calibrated, the cluster
ZAMS-fitting method may not be the best way to derive the zero point of
the Cepheid PL relation. On the other hand,
the determination of the PL zero point,
and thus of the distance to the LMC, from HIPPARCOS parallaxes of Galactic
Cepheids \citep{FC97}, while of course very attractive as an independent,
geometrical method, has proven to be a difficult subject, due to the
crucial importance of the correct treatment of biases in these very
low signal-to-noise data. The $\pm8$ percent (statistical error only)
uncertainty of the LMC distance obtained with this method \citep{pont99}
is still too large for a galaxy which is our closest neighbor in space and
serves as a principal reference object for the extragalactic distance
scale.
      For a long time, the best alternative to derive distances of Cepheid
variables has been the Baade-Wesselink method \citep{wesselink46}
which utilizes the light-, color- and radial velocity variations of a
Cepheid to infer its radius and distance in a way which is completely
independent of other astrophysical distance scales. Over the years, the
classic approach used by Wesselink has been improved in many respects
(for a good review, see \citet{gautschy87}).  Perhaps the single most
important improvement of the method has been the shift to infrared
wavelengths, where the dependence of the surface brightness of a Cepheid
on atmospheric parameters like gravity and microturbulence \citep{LS95},
and on metallicity \citep{BG89} becomes very small as compared to
optical wavelengths. This has opened up the possibility to derive the
distances to Cepheid variables with an accuracy of $\approx 4$~percent
when the $V$, $V-K$ or $K$, $J-K$ magnitude/color combinations are used
to derive the Cepheid surface brightnesses in the $V$ or $K$ band.
The surface brightnesses yield the angular diameters once the relation between
surface brightness and angular diameter has been properly calibrated. Such
an empirical calibration was first presented by \citet{welch94}
using precise, interferometrically measured angular diameters of giants
and supergiants in the Cepheid color range, and has more recently been
improved and extended by \citet{FG97}. These authors
have shown \citep{GFG97,GFG98} that the infrared
surface brightness (ISB) technique, as calibrated in \citet{FG97},
yields Cepheid distances from either of its versions (using $V$, $V-K$
or $K$, $J-K$) which are accurate to about 4 percent if the underlying
observational data are of excellent quality and the amplitude of the
color variation exceeds $\approx 0.25$~mag. In $V-K$, this is the case
for most of even the shortest period Cepheids pulsating in the fundamental
mode. The ISB technique is therefore an excellent tool to derive direct,
accurate distances to Cepheid variables which are likely independent
of the stellar metallicity. As an added benefit, they are also largely
independent of the absorption corrections applied in the analyses
\citep{welch94,GFG97}. These latter properties make the method very
attractive for an application to extragalactic Cepheids.  Here, the
obvious target to start with is the LMC for which the necessary data can
be acquired with small-to medium-aperture telescopes. The ideal targets
in the LMC for the potentially most accurate distance determination with
Cepheids are the young, rich clusters.  Due to the right combination of
turnoff mass, richness and metallicity, a number of these clusters have
been found to contain considerable numbers of Cepheids \citep{mateo92}
which allow a distance determination to these clusters, and thus to the
LMC barycenter, more accurate than that achievable for individual field
variables. It is therefore reasonable to expect that an application of
the  ISB technique to Cepheid-rich LMC globular clusters will produce
the as yet most accurate distance determination to the LMC based on
Cepheid variables, and allow the determination of the absolute PL zero
point (in a given photometric band) with a smaller uncertainty than other
methods. In the construction of the extragalactic distance ladder, and in
order to take full advantage of the HST Key Project Cepheid observations
in other galaxies, this is a crucial step, especially in view of the
large dispersion among LMC distance moduli as currently obtained from
a variety of objects and methods (for a comprehensive review of this
subject, see \citet{walker99}).
     We start our work on Cepheids in LMC globular clusters with 
NGC~1866. More than 20 Cepheids have been detected in this cluster
(\citet{welch91,Storm88}; and references therein), making it the
most Cepheid-rich cluster known in the LMC.  Previous work on the variable
stars in NGC~1866, and particularly on its Cepheids, has been reviewed
in \citet{welch91} who were also the first to collect radial velocity
data on a number of Cepheids in the cluster field. Recently, our group
has reported results on a first Cepheid variable in NGC~1866, HV~12198,
from our current study using the ISB technique \citep{gieren2000}. In
that paper, we also reviewed more recent ($>1991$) work on NGC~1866
Cepheids. The purpose of this paper is to present the new optical
CCD photometry we have obtained for a number of NGC~1866 Cepheids,
including HV~12198. A follow-up paper \citep{Storm2000} will present
infrared photometry and new radial velocity observations for these
same variables. This material will then be used to discuss cluster
membership and binarity for the Cepheids in the field of NGC~1866, and
to derive their distances and radii with the ISB method. 
We emphasize that one of the
important consequences of many Cepheids being in the same cluster is a
sanity check that the actual precision of the technique can be assessed
(without the usual fallback explanations of differential reddening,
distance modulus, and/or metallicity).
While we shall also study the physical
and evolutionary properties of the Cepheids in NGC~1866, our principal
goal is a truly accurate distance determination to the LMC, which will
be further improved by distance determinations to other clusters we are
currently working on, notably NGC~2031 and NGC~2136/37.

\section{Photometry of selected Cepheids in NGC1866}

\begin{figure*}[htp]
\epsfigx{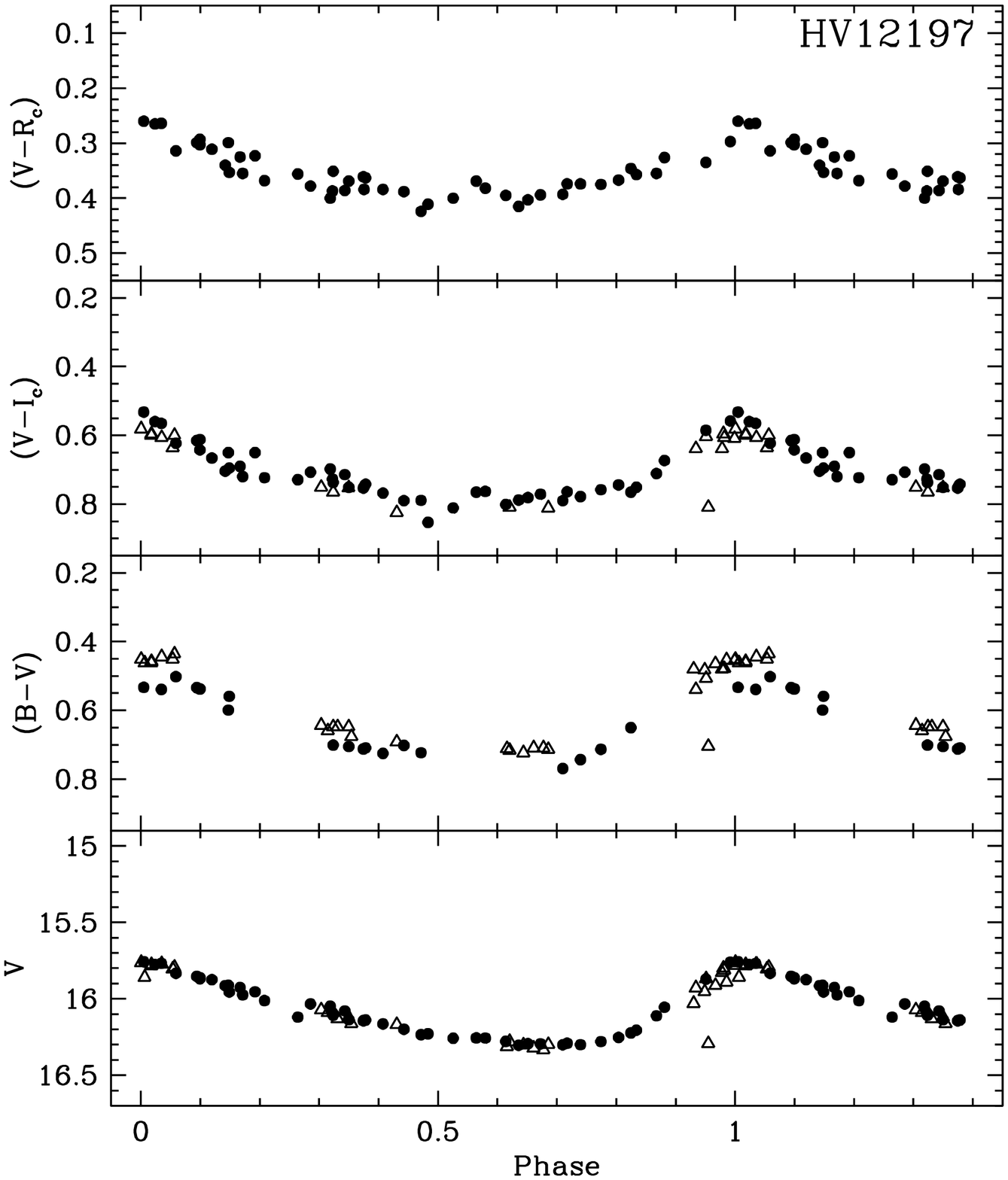}
\figcaption[fig1.eps]{Light and color curves for the Cepheid variable
HV~12197 in the young, rich LMC cluster NGC~1866. Phases were calculated
with the periods given in Table 8. Filled circles: new CCD data reported
in this paper. Open triangles: literature CCD data from Walker (1987),
and Welch et. al (1991).
\label{hv12197lc.fig}}
\end{figure*}

\begin{figure*}[htp]
\epsfigx{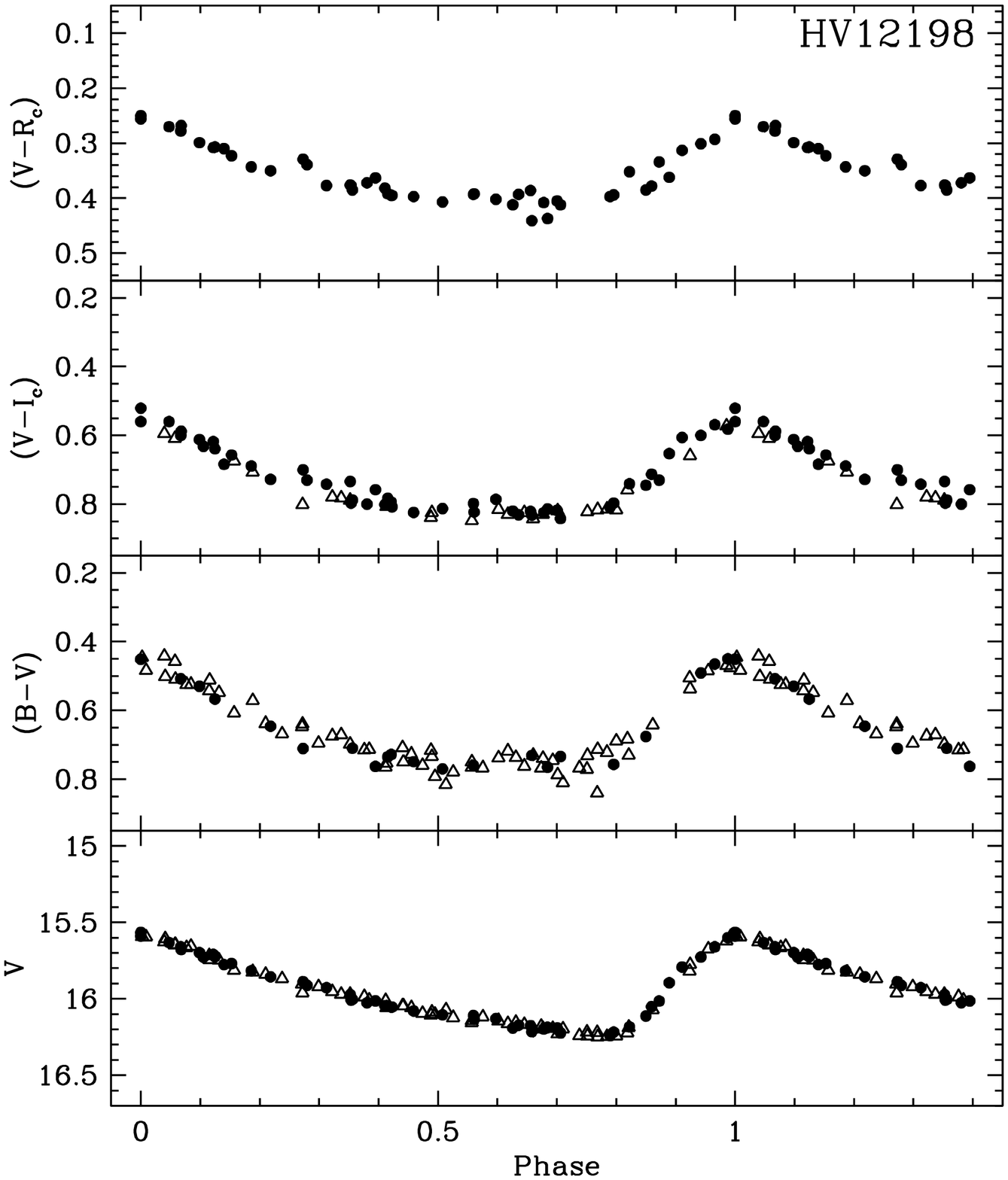}
\figcaption[fig2.eps]{Light and color curves for the Cepheid variable
HV~12198. The points have been marked as in Fig.\ref{hv12197lc.fig}.
\label{hv12198lc.fig}}
\end{figure*}

\begin{figure*}[htp]
\epsfigx{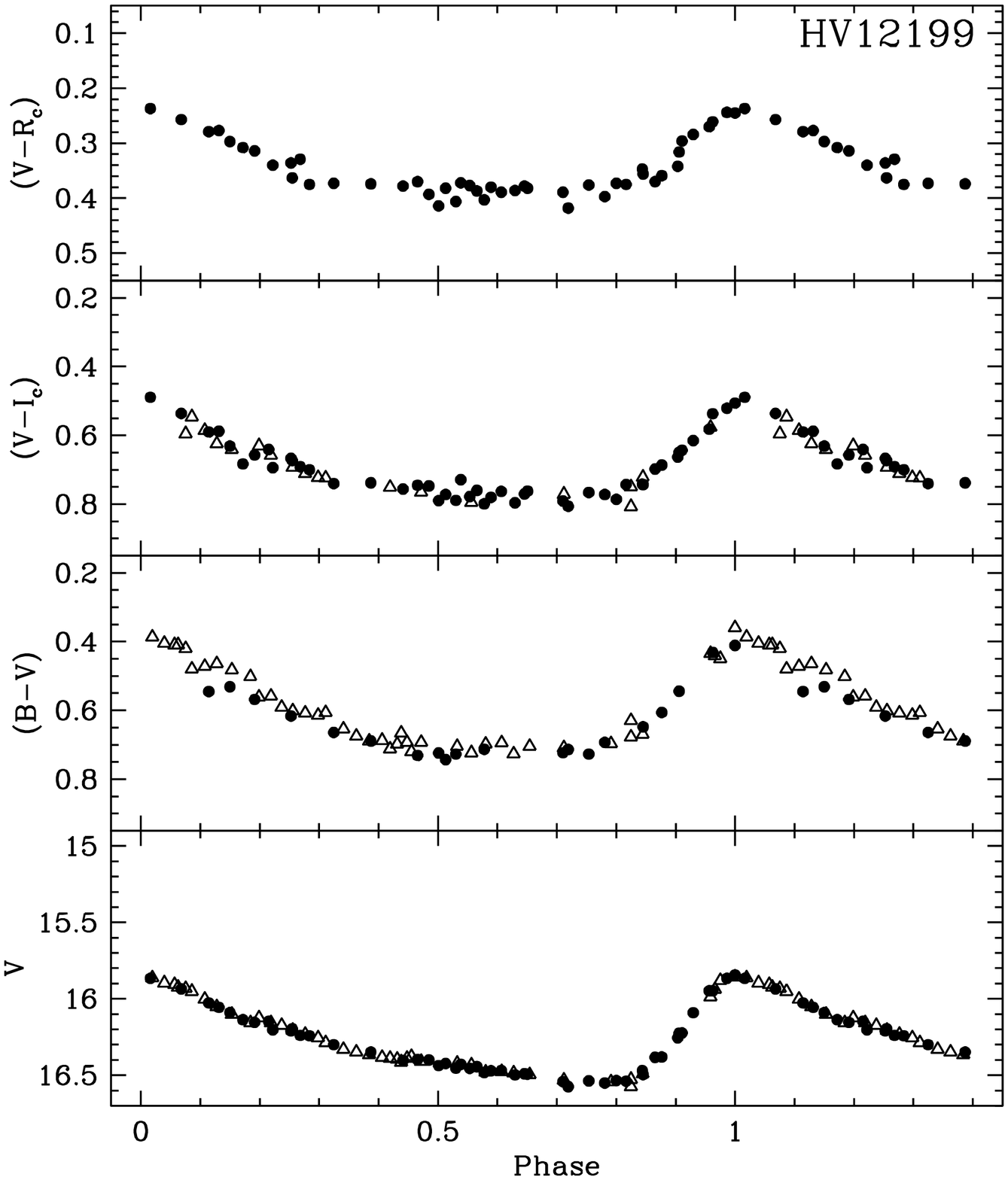}
\figcaption[fig3.eps]{Light and color curves for the Cepheid variable
HV~12199. The points have been marked as in Fig.\ref{hv12197lc.fig}.
\label{hv12199lc.fig}}
\end{figure*}

\begin{figure*}[htp]
\epsfigx{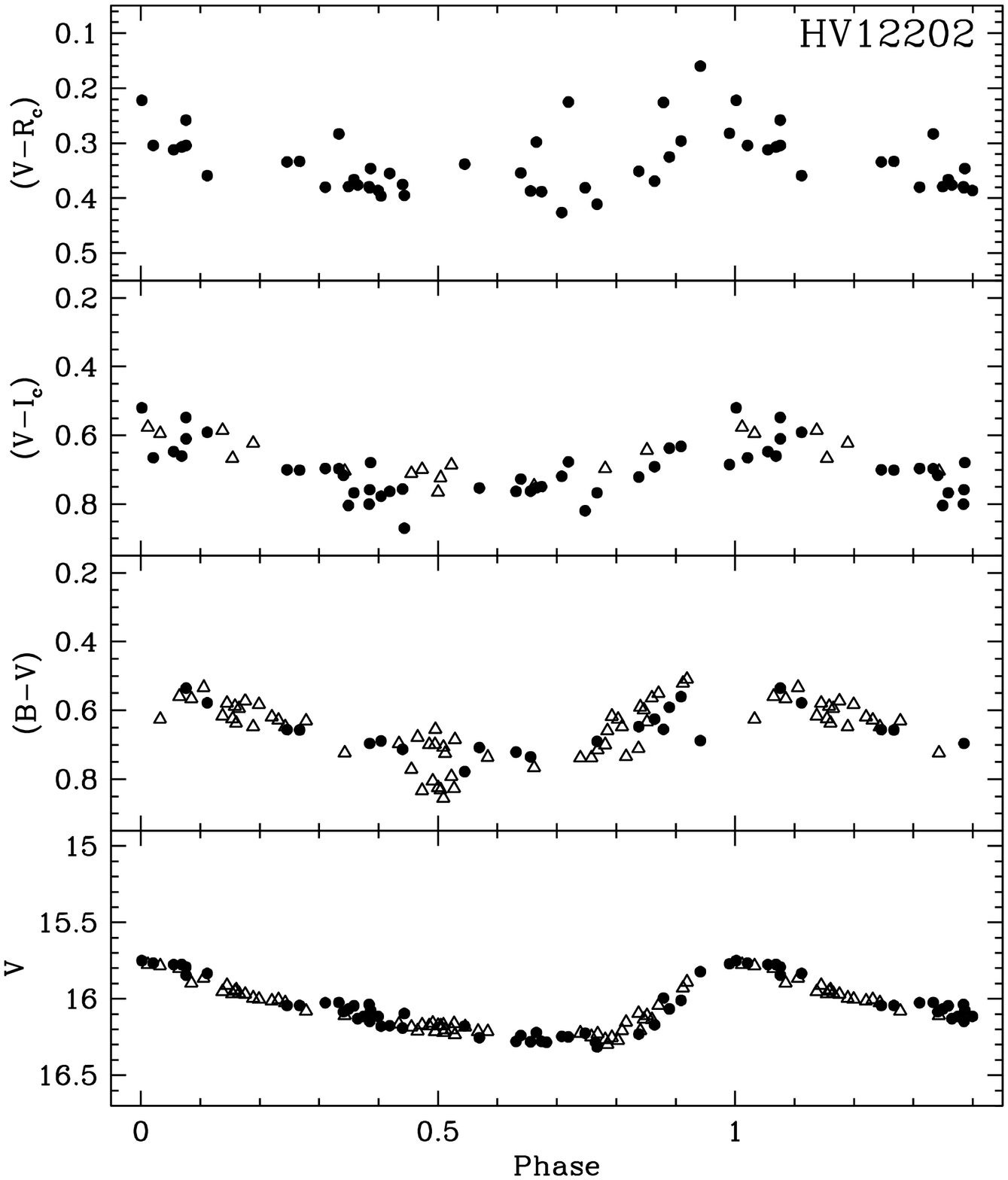}
\figcaption[fig4.eps]{Light and color curves for the Cepheid variable
HV~12202. The points have been marked as in Fig.\ref{hv12197lc.fig}.
\label{hv12202lc.fig}}
\end{figure*}

\begin{figure*}[htp]
\epsfigx{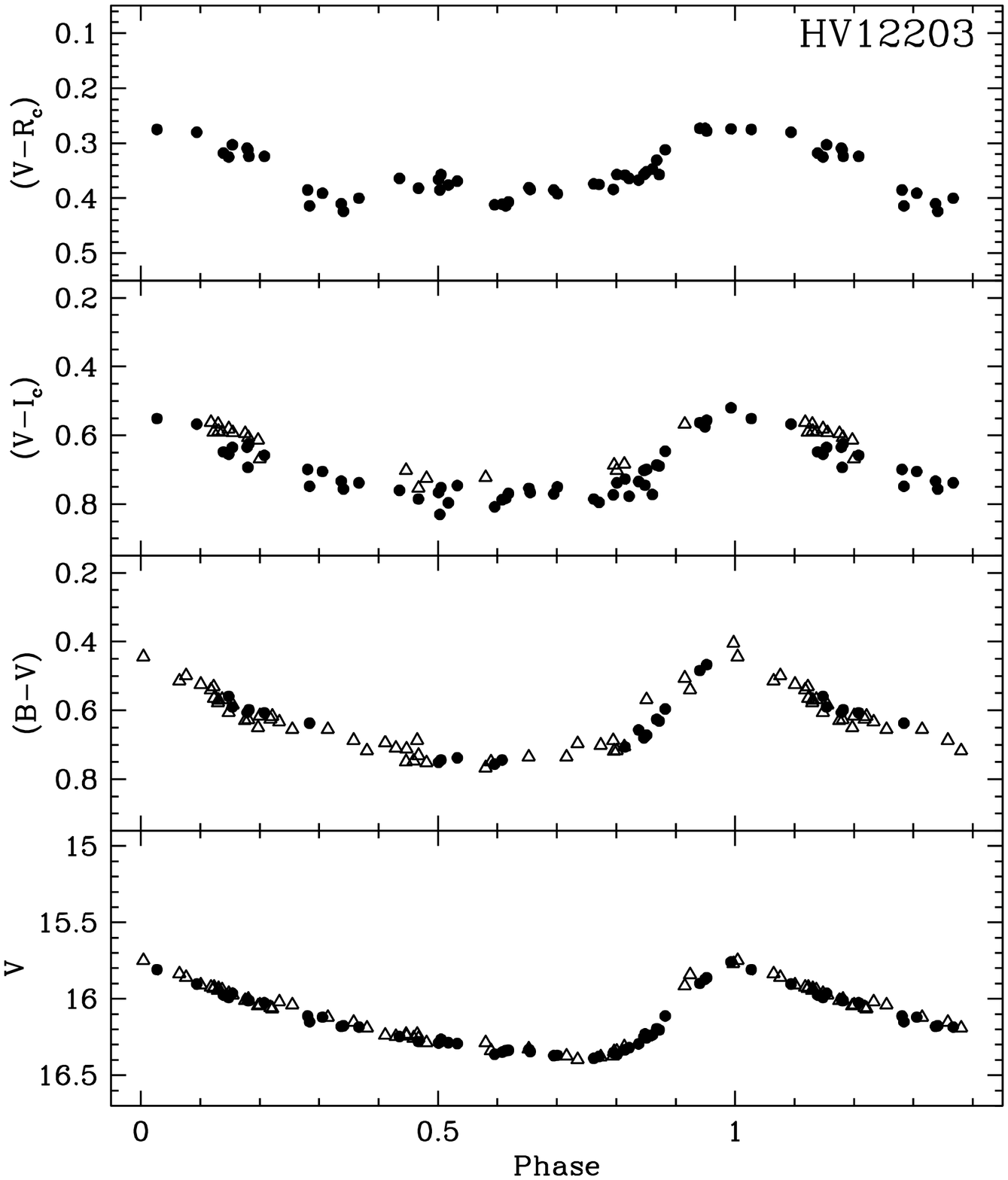}
\figcaption[fig5.eps]{Light and color curves for the Cepheid variable
HV~12203. The points have been marked as in Fig.\ref{hv12197lc.fig}.
\label{hv12203lc.fig}}
\end{figure*}

\begin{figure*}[htp]
\epsfigx{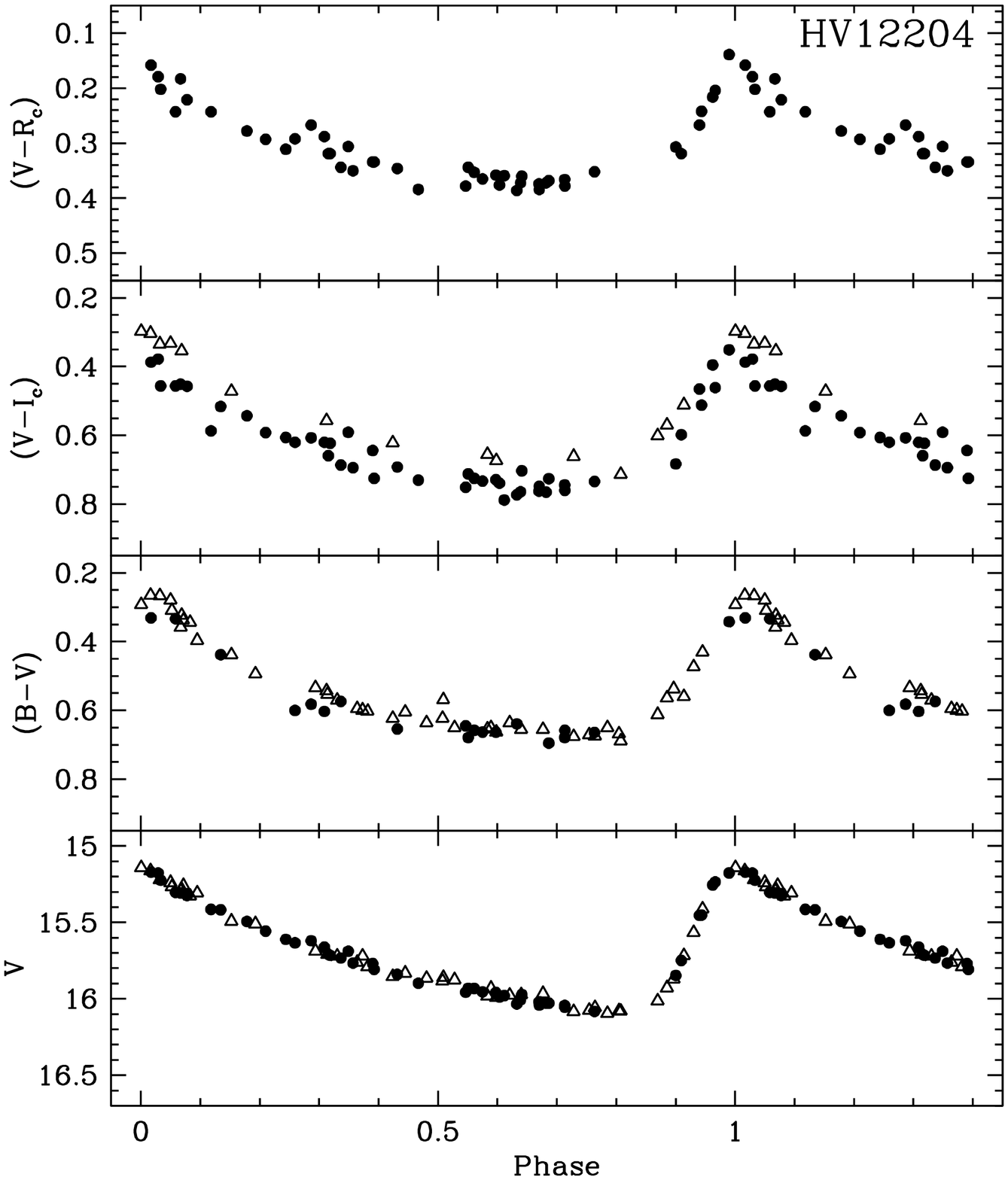}
\figcaption[fig6.eps]{Light and color curves for the Cepheid variable
HV~12204. The points have been marked as in Fig.\ref{hv12197lc.fig}.
\label{hv12204lc.fig}}
\end{figure*}

\begin{figure*}[htp]
\epsfigx{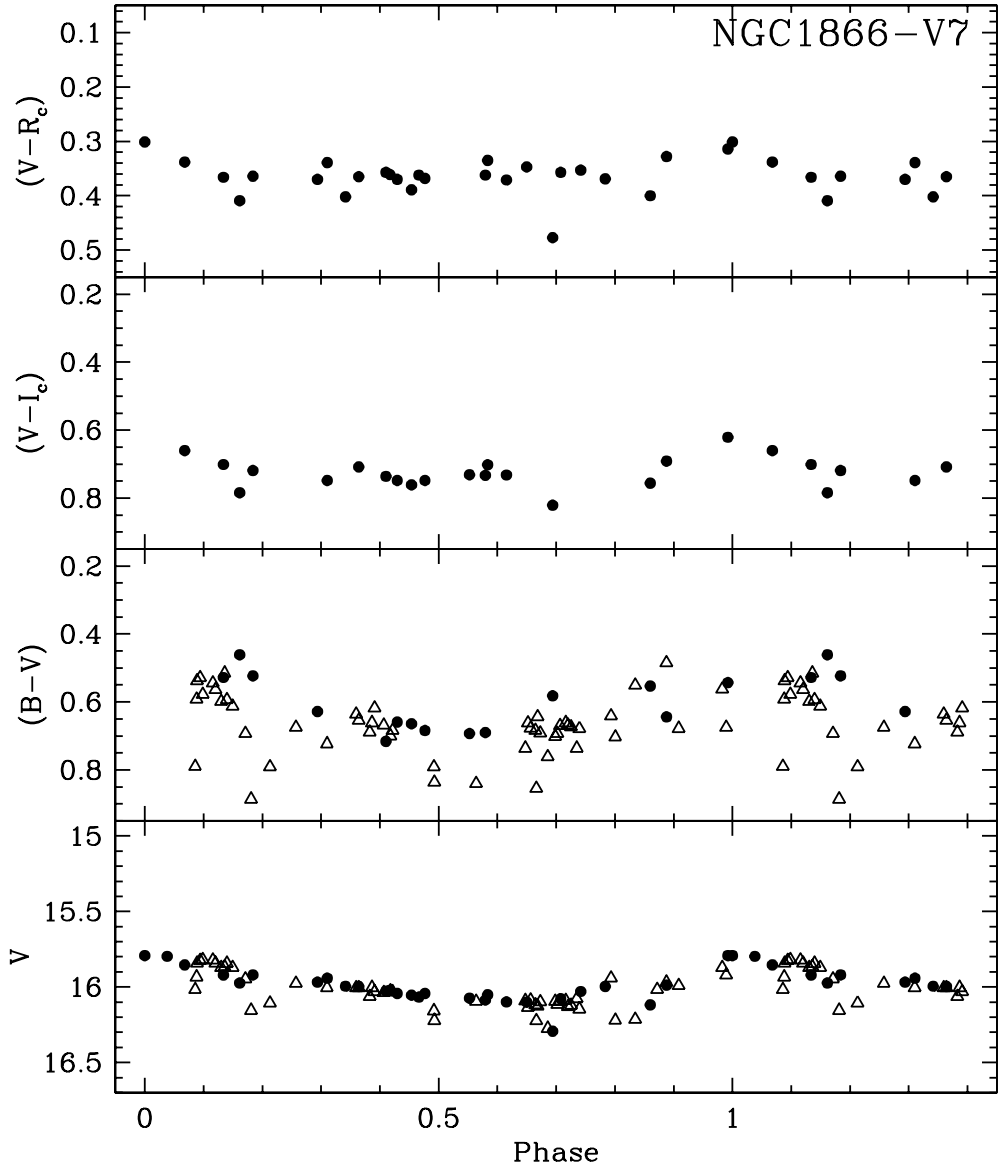}
\figcaption[fig7.eps]{Light and color curves for the Cepheid variable
NGC1866-V7. The points have been marked as in Fig.\ref{hv12197lc.fig}.
\label{V7lc.fig}}
\end{figure*}

\begin{figure*}[htp]
\epsfigx{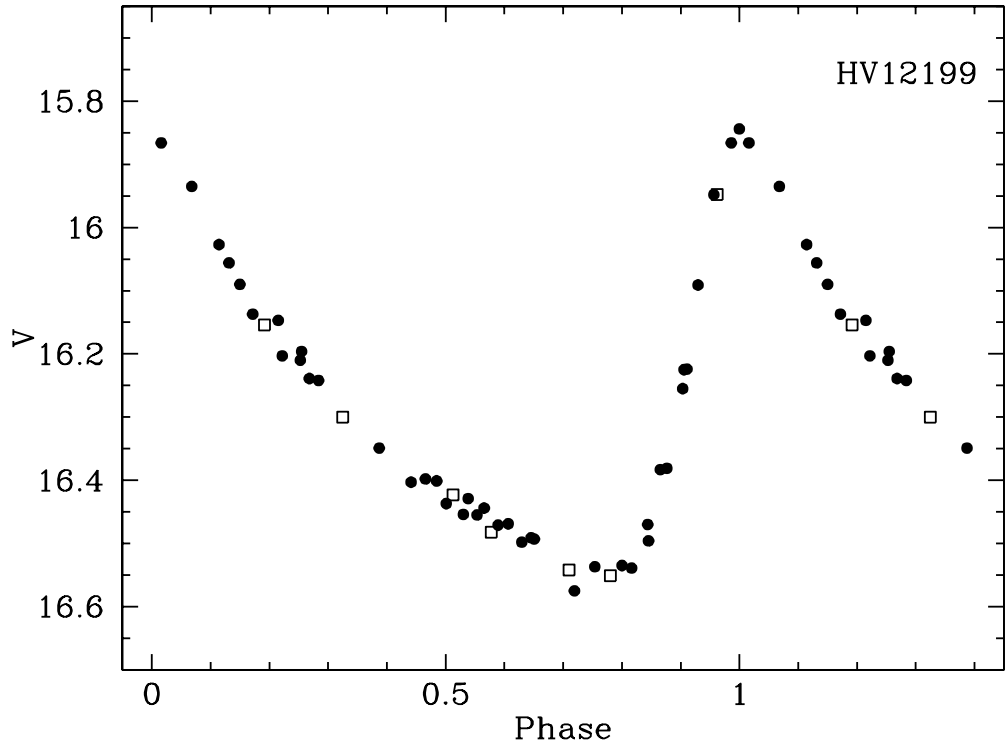}
\figcaption[fig8.eps]{$V$ light curve of HV~12199, from our new CCD data.
  Filled circles: Las Campanas data, reduced with DoPHOT. Open boxes:
  CTIO data, reduced with DAOPHOT. The data sets are consistent at
  the 0.01~mag level. This holds also for the data in the B,R and I bands
  (not shown here). \label{hv12199comp.fig}}
\end{figure*}

     CCD images in $BVRI$ (Cousins) filters of NGC~1866 were acquired
     during fours
runs between November 1994 and January 1996. The first sequence of
images was obtained using the CFCCD instrument on the 0.9~m telescope
of CTIO which has a pixel scale of 0.58~arcsec/pixel,
while the remainder of the data was obtained on the Swope 1.0~m
reflector of Las Campanas Observatory, using the 1024 x 1024 TEK2 CCD
which has a
pixel scale of 0.6 arcsec/pixel and a field of view of 10 sq. arcmin.
Most of the observing nights were photometric. Integration times of
240~sec in all of the filters produced a S/N for the Cepheids always
high enough for photometry at the 0.01-0.02 mag level of accuracy (see
below). On all the photometric nights, we observed large numbers of
photometric standard stars from the list of \citet{Landolt92} to tie our
observations of the Cepheids to the standard $BVRI$ system. From some 20
local comparison stars on our frames which were checked thoroughly for
photometric constancy we established a final local comparison by taking a
weighted sum of the intensities of this set of stars, in any of the four
filters. The weighting was proportional to the intensities of the local
comparison stars which ensured that stars with higher signal to noise
ratios got a higher weight. All the photometry was done differentially
with respect to this local reference system. Using this differential
photometry procedure, we could obtain reliable Cepheid magnitudes on a
number of non-photometric nights.
     Once the images were bias-subtracted and flatfielded, point spread
function-photometry was done on them using the DAOPHOT \citep{stetson87}
and DoPHOT \citep{SMS93} reduction  packages for photometry in crowded
fields. We chose to do the automatic finding and photometry down to the
faintest objects, and then to identify the Cepheids on the output lists.
DAOPHOT was used by one of us (TJM) to reduce the 1994 CTIO
images while DoPHOT was used to reduce all the images obtained at Las
Campanas. This procedure yielded an independent check on the accuracy
of our photometry, on an absolute scale, and the light curve plots
(Figs. \ref{hv12197lc.fig}-\ref{V7lc.fig}) of the Cepheids show that there
is excellent consistency (at the ~0.01 mag level) between the magnitudes,
in all filters, derived with both software packages. An example is shown
in Fig. \ref{hv12199comp.fig} for the $V$-band magnitudes of the variable
HV~12199. We did photometry on the "classical" Cepheid variables HV~12197,
HV~12198, HV~12199, HV~12202, HV~12203 and HV~12204 in the NGC~1866 field
(for a finder chart of these variables, see \citet{AT67,Storm88,welch91})
which i) seemed to be sufficiently uncrowded to allow photometry accurate
enough for the distance and radius determinations we are going to perform
with these data, and ii) have light- and color amplitudes large enough
to permit a successful use of at least the $V$, $V-K$ version of the ISB
technique. The individual photometric observations for these six variables
are listed in Tables \ref{HV12197phot.tab}-\ref{HV12204phot.tab}.
The first column gives the Heliocentric Julian Dates for the mid-points
of a $BVRI$ (or $VRI$) sequence, column 2 the phases from maximum light
(see below), and the other columns give the calibrated $V$ magnitudes
and $B-V$, $V-I$ and $V-R$ color indices, each with their respective
standard deviations. To check on the capabilities of the DoPHOT routine
to do photometry in very crowded fields, we also performed photometry
on the Cepheid variable V7 (see \citet{Storm88}) which is located
in the central region of the cluster. These data are given in Table
\ref{V7phot.tab}. As expected, the light curve (Fig. \ref{V7lc.fig})
is much noisier than that for the other variables, ruling this variable
(and the other Cepheids close to the center of NGC~1866) out for a
distance determination with the ISB technique.
     We used our new $V$ band data together with the literature $V$
data reported by \citet{walker87} and \citet{welch91} to derive improved
values for the pulsation periods of the Cepheids. These values differ
only slightly from those given and discussed in \citet{Cote91}; the
agreement is good to 2 - 5$\times 10^{-5}$~days. 

  We can confirm the variable period of HV~12198 as found by
\citet{Cote91}, but the period change is so small that there is no
significant difference between their period and ours, and consequently 
there is no significant effect on the phasing between the data presented
here and those of \citet{Cote91}.

The improved periods
are listed in Table \ref{ephem.tab}, together with modern epochs of
maximum light (in $V$) of the variables derived from our new data. 
For the Cepheid V7, the uncertainty
of the period is much larger as for the other variables since we only
have data from a few epochs and therefore run into aliasing problems.
The number of digits given in Table \ref{ephem.tab} is still significant,
however, if we assume that we have adopted the correct period peak.
In Figs. \ref{hv12197lc.fig}-\ref{V7lc.fig},
we show the light- and color curves of the Cepheids, folded on the new
periods derived in this paper. 
Overplotted on the new data are the $BVI$ data of \citet{walker87},
and the $BV$ data of \citet{welch91}.  The new data are more complete
and slightly more accurate than the previous data sets. It is seen that
for five of the Cepheid variables in NGC~1866, the composite $V$ light
curves are now of excellent quality, exhibiting complete phase coverage
and very low scatter. Only the two Cepheids closest to the cluster
center, HV~12202 and V7, show enhanced scatter in their light curves
which is clearly a result of increased contamination in the photometry
due to severe crowding. For all of the variables, the agreement of the
various data sets in $V$ is excellent. Offsets typically amounting to
0.05 mag between different data sets are visible, however, in the $B$
and $I$ bands. Our new $B$ band data have a tendency to be somewhat redder
than the Walker and Welch et al. data, which among themselves also show
minor systematic deviations as already discussed by \citet{welch91}. The
$V-I$ data show the same trend, but for a few stars, like HV~12198 and
HV~12199, the agreement between our and the Walker data is very good. It
is likely that these small zero point differences have their origin in
different amounts of contamination by nearby blue and red stars due to
different pixel scales in the CCDs used by the different authors, or are
simply a consequence of small systematic errors in the transformation
to the standard system. In this context, we are very confident that
our magnitudes match the $BVRI$ standard system very well. 
We find formal rms uncertainties for the transformation to
the standard system of $\pm 0.014$ in $B$, $\pm 0.015$ in $V$, 
$\pm 0.018$ in $R$, and $\pm 0.017$ in $I$, and these small
uncertainties are supported by the excellent
agreement of the CTIO and LCO data sets in all of the filters which were
reduced in a completely independent way.

     For the $V$, $V-K$ infrared surface brightness distance and radius
determination for which these data were principally obtained, it is
comforting to see that the $V$ light curves of five variables are now
very accurately defined, and that in this band there are evidently no
zero point problems in the photometry. These five Cepheids are therefore
excellent candidates for accurate distance determinations once their
$K$ light and radial velocity curves become available. We are currently
working on this.

\acknowledgements

WPG and MG gratefully acknowledge substantial amounts of observing time
granted to this project by the Las Campanas Observatory. They appreciate
the expert support of the LCO staff which makes observing at Las Campanas
a very pleasant experience. WPG also acknowledges financial support for
this project received by CONICYT through Fondecyt grant 1971076,
and by the Universidad de Concepci\'on through research grant No.
97.011.021-1.0.  TJM and TGB gratefully acknowledge
observing time at CTIO and the strong
support received from the staff. They acknowledge financial support from
NATO grant 900494 and National Science Foundations grants AST 92-21595
(TJM), AST 95-28134 (TJM), AST 92-21405 (TGB), and AST 95-28372 (TGB).


\clearpage



\clearpage

\clearpage

\clearpage

\clearpage

\clearpage

\clearpage

\clearpage

\clearpage

\begin{deluxetable}{rrrrrrrrrr}
\tabletypesize{\scriptsize}
\tablecaption{Photometry for HV12197. \label{HV12197phot.tab}}
\tablewidth{0pt}
\tablehead{
\colhead{HJD-2400000} & \colhead{Phase}   & 
\colhead{$V$}   & \colhead{$\sigma_V$} &
\colhead{$(B-V)$}   & \colhead{$\sigma_{B-V}$} &
\colhead{$(V-I)$}   & \colhead{$\sigma_{V-I}$} &
\colhead{$(V-R)$}   & \colhead{$\sigma_{V-R}$} 
}
\startdata
 49651.841 & 0.4427 & 16.199 & 0.007 &  0.702 & 0.011 &  0.790 & 0.012 &  0.388 & 0.012 \\ 
 49668.761 & 0.8246 & 16.223 & 0.005 &  0.650 & 0.008 &  0.765 & 0.011 &  0.346 & 0.009 \\ 
 49669.609 & 0.0943 & 15.853 & 0.007 &  0.534 & 0.011 &  0.615 & 0.013 &  0.299 & 0.011 \\ 
 49669.780 & 0.1488 & 15.955 & 0.006 &  0.559 & 0.009 &  0.695 & 0.012 &  0.353 & 0.010 \\ 
 49670.795 & 0.4715 & 16.235 & 0.007 &  0.723 & 0.012 &  0.789 & 0.012 &  0.424 & 0.013 \\ 
 49672.769 & 0.0995 & 15.869 & 0.006 &  0.537 & 0.009 &  0.642 & 0.012 &  0.293 & 0.010 \\ 
 49713.788 & 0.1471 & 15.912 & 0.007 &  0.599 & 0.011 &  0.650 & 0.012 &  0.299 & 0.012 \\ 
 49715.760 & 0.7742 & 16.280 & 0.009 &  0.713 & 0.013 &  0.758 & 0.012 &  0.375 & 0.012 \\ 
 49716.782 & 0.0994 & 15.861 & 0.006 &  0.538 & 0.008 &  0.612 & 0.008 &  0.303 & 0.008 \\ 
 49717.658 & 0.3781 & 16.139 & 0.006 &  0.709 & 0.009 &  0.742 & 0.008 &  0.363 & 0.008 \\ 
 49717.750 & 0.4073 & 16.164 & 0.006 &  0.725 & 0.009 &  0.768 & 0.008 &  0.384 & 0.008 \\ 
 49718.701 & 0.7099 & 16.301 & 0.018 &  0.769 & 0.021 &  0.790 & 0.022 &  0.393 & 0.023 \\ 
 49718.795 & 0.7397 & 16.300 & 0.007 &  0.743 & 0.011 &  0.778 & 0.010 &  0.374 & 0.011 \\ 
 49719.628 & 0.0048 & 15.757 & 0.014 &  0.533 & 0.018 &  0.532 & 0.016 &  0.260 & 0.018 \\ 
 49719.722 & 0.0346 & 15.768 & 0.017 &  0.539 & 0.018 &  0.565 & 0.018 &  0.264 & 0.022 \\ 
 49719.799 & 0.0589 & 15.833 & 0.008 &  0.502 & 0.013 &  0.623 & 0.011 &  0.314 & 0.011 \\ 
 49720.631 & 0.3236 & 16.108 & 0.011 &  0.701 & 0.014 &  0.737 & 0.015 &  0.351 & 0.016 \\ 
 49720.713 & 0.3499 & 16.137 & 0.013 &  0.705 & 0.017 &  0.751 & 0.017 &  0.369 & 0.016 \\ 
 49720.790 & 0.3744 & 16.144 & 0.008 &  0.712 & 0.011 &  0.753 & 0.011 &  0.361 & 0.010 \\ 
 50069.590 & 0.3225 & 16.092 & 0.010 &   ---  &  ---  &  0.727 & 0.011 &  0.387 & 0.012 \\ 
 50069.757 & 0.3756 & 16.141 & 0.009 &   ---  &  ---  &  0.749 & 0.010 &  0.384 & 0.013 \\ 
 50070.575 & 0.6358 & 16.304 & 0.013 &   ---  &  ---  &  0.788 & 0.016 &  0.415 & 0.017 \\ 
 50070.690 & 0.6724 & 16.295 & 0.007 &   ---  &  ---  &  0.771 & 0.009 &  0.394 & 0.009 \\ 
 50070.831 & 0.7173 & 16.291 & 0.008 &   ---  &  ---  &  0.764 & 0.010 &  0.374 & 0.011 \\ 
 50071.565 & 0.9508 & 15.871 & 0.011 &   ---  &  ---  &  0.585 & 0.014 &  0.335 & 0.015 \\ 
 50071.694 & 0.9919 & 15.760 & 0.007 &   ---  &  ---  &  0.558 & 0.011 &  0.297 & 0.011 \\ 
 50071.795 & 0.0239 & 15.776 & 0.010 &   ---  &  ---  &  0.560 & 0.012 &  0.265 & 0.013 \\ 
 50072.617 & 0.2855 & 16.033 & 0.005 &   ---  &  ---  &  0.707 & 0.007 &  0.378 & 0.008 \\ 
 50072.722 & 0.3188 & 16.049 & 0.009 &   ---  &  ---  &  0.698 & 0.013 &  0.400 & 0.012 \\ 
 50072.799 & 0.3431 & 16.079 & 0.009 &   ---  &  ---  &  0.714 & 0.011 &  0.386 & 0.013 \\ 
 50073.542 & 0.5797 & 16.257 & 0.010 &   ---  &  ---  &  0.763 & 0.013 &  0.382 & 0.015 \\ 
 50073.651 & 0.6141 & 16.278 & 0.010 &   ---  &  ---  &  0.801 & 0.012 &  0.395 & 0.013 \\ 
 50073.767 & 0.6513 & 16.294 & 0.009 &   ---  &  ---  &  0.781 & 0.010 &  0.403 & 0.011 \\ 
 50100.553 & 0.1715 & 15.974 & 0.008 &   ---  &  ---  &  0.720 & 0.013 &  0.355 & 0.015 \\ 
 50100.669 & 0.2081 & 16.012 & 0.014 &   ---  &  ---  &  0.723 & 0.024 &  0.368 & 0.021 \\ 
 50100.845 & 0.2642 & 16.120 & 0.007 &   ---  &  ---  &  0.729 & 0.010 &  0.356 & 0.009 \\ 
 50101.534 & 0.4833 & 16.229 & 0.021 &   ---  &  ---  &  0.853 & 0.048 &  0.411 & 0.026 \\ 
 50101.667 & 0.5256 & 16.258 & 0.009 &   ---  &  ---  &  0.811 & 0.030 &  0.400 & 0.016 \\ 
 50101.788 & 0.5643 & 16.256 & 0.007 &   ---  &  ---  &  0.765 & 0.017 &  0.369 & 0.010 \\ 
 50102.542 & 0.8039 & 16.253 & 0.015 &   ---  &  ---  &  0.744 & 0.024 &  0.367 & 0.019 \\ 
 50102.636 & 0.8339 & 16.205 & 0.017 &   ---  &  ---  &  0.751 & 0.023 &  0.357 & 0.023 \\ 
 50102.742 & 0.8677 & 16.111 & 0.008 &   ---  &  ---  &  0.711 & 0.011 &  0.355 & 0.012 \\ 
 50102.785 & 0.8812 & 16.055 & 0.011 &   ---  &  ---  &  0.673 & 0.014 &  0.326 & 0.015 \\ 
 50103.534 & 0.1195 & 15.875 & 0.011 &   ---  &  ---  &  0.666 & 0.014 &  0.311 & 0.016 \\ 
 50103.605 & 0.1421 & 15.914 & 0.009 &   ---  &  ---  &  0.704 & 0.015 &  0.340 & 0.017 \\ 
 50103.683 & 0.1670 & 15.926 & 0.009 &   ---  &  ---  &  0.690 & 0.015 &  0.325 & 0.011 \\ 
 50103.762 & 0.1922 & 15.954 & 0.010 &   ---  &  ---  &  0.650 & 0.020 &  0.323 & 0.015 \\ 
\enddata

\end{deluxetable}

\clearpage

\begin{deluxetable}{rrrrrrrrrr}
\tabletypesize{\scriptsize}
\tablecaption{Photometry for HV12198. \label{HV12198phot.tab}}
\tablewidth{0pt}
\tablehead{
\colhead{HJD-2400000} & \colhead{Phase}   & 
\colhead{$V$}   & \colhead{$\sigma_V$} &
\colhead{$(B-V)$}   & \colhead{$\sigma_{B-V}$} &
\colhead{$(V-I)$}   & \colhead{$\sigma_{V-I}$} &
\colhead{$(V-R)$}   & \colhead{$\sigma_{V-R}$} 
}
\startdata
 49651.841 & 0.4154 & 16.047 & 0.006 &  0.735 & 0.010 &  0.783 & 0.013 &  0.392 & 0.010 \\ 
 49668.761 & 0.2182 & 15.856 & 0.004 &  0.646 & 0.006 &  0.728 & 0.009 &  0.350 & 0.007 \\ 
 49669.609 & 0.4589 & 16.080 & 0.006 &  0.749 & 0.010 &  0.824 & 0.010 &  0.397 & 0.010 \\ 
 49669.780 & 0.5076 & 16.105 & 0.005 &  0.770 & 0.009 &  0.813 & 0.009 &  0.407 & 0.009 \\ 
 49670.795 & 0.7956 & 16.218 & 0.006 &  0.757 & 0.009 &  0.797 & 0.012 &  0.394 & 0.010 \\ 
 49671.751 & 0.0672 & 15.660 & 0.004 &  0.508 & 0.006 &  0.600 & 0.009 &  0.278 & 0.007 \\ 
 49672.769 & 0.3560 & 16.004 & 0.005 &  0.710 & 0.008 &  0.788 & 0.010 &  0.385 & 0.009 \\ 
 49713.788 & 0.0000 & 15.589 & 0.009 &  0.451 & 0.012 &  0.560 & 0.013 &  0.256 & 0.014 \\ 
 49714.750 & 0.2730 & 15.888 & 0.016 &  0.711 & 0.020 &  0.700 & 0.021 &  0.329 & 0.025 \\ 
 49715.760 & 0.5596 & 16.110 & 0.010 &  0.760 & 0.015 &  0.798 & 0.013 &  0.393 & 0.014 \\ 
 49716.782 & 0.8498 & 16.113 & 0.007 &  0.676 & 0.011 &  0.745 & 0.010 &  0.385 & 0.009 \\ 
 49717.658 & 0.0986 & 15.698 & 0.005 &  0.530 & 0.007 &  0.612 & 0.008 &  0.299 & 0.007 \\ 
 49717.750 & 0.1246 & 15.725 & 0.005 &  0.567 & 0.008 &  0.639 & 0.007 &  0.307 & 0.008 \\ 
 49718.701 & 0.3947 & 16.014 & 0.021 &  0.763 & 0.023 &  0.758 & 0.024 &  0.363 & 0.026 \\ 
 49718.795 & 0.4213 & 16.054 & 0.008 &  0.728 & 0.012 &  0.794 & 0.011 &  0.395 & 0.011 \\ 
 49719.628 & 0.6579 & 16.216 & 0.015 &  0.731 & 0.018 &  0.832 & 0.017 &  0.441 & 0.019 \\ 
 49719.722 & 0.6844 & 16.186 & 0.027 &  0.765 & 0.029 &  0.814 & 0.028 &  0.437 & 0.031 \\ 
 49719.799 & 0.7061 & 16.223 & 0.011 &  0.734 & 0.015 &  0.842 & 0.014 &  0.412 & 0.015 \\ 
 49720.631 & 0.9423 & 15.727 & 0.013 &  0.491 & 0.015 &  0.600 & 0.016 &  0.301 & 0.016 \\ 
 49720.713 & 0.9658 & 15.660 & 0.013 &  0.465 & 0.015 &  0.569 & 0.016 &  0.293 & 0.016 \\ 
 49720.790 & 0.9876 & 15.600 & 0.008 &  0.450 & 0.011 &  0.582 & 0.011 &   ---  &  ---  \\ 
 50069.590 & 0.0000 & 15.565 & 0.006 &   ---  &  ---  &  0.521 & 0.008 &  0.250 & 0.008 \\ 
 50069.757 & 0.0474 & 15.632 & 0.007 &   ---  &  ---  &  0.560 & 0.009 &  0.270 & 0.009 \\ 
 50070.575 & 0.2796 & 15.912 & 0.008 &   ---  &  ---  &  0.730 & 0.009 &  0.339 & 0.012 \\ 
 50070.690 & 0.3123 & 15.927 & 0.006 &   ---  &  ---  &  0.742 & 0.008 &  0.377 & 0.008 \\ 
 50070.831 & 0.3524 & 15.973 & 0.008 &   ---  &  ---  &  0.734 & 0.009 &  0.376 & 0.010 \\ 
 50071.565 & 0.5607 & 16.133 & 0.010 &   ---  &  ---  &  0.823 & 0.011 &  0.392 & 0.013 \\ 
 50071.694 & 0.5974 & 16.130 & 0.007 &   ---  &  ---  &  0.786 & 0.010 &  0.402 & 0.009 \\ 
 50071.795 & 0.6260 & 16.191 & 0.006 &   ---  &  ---  &  0.820 & 0.008 &  0.412 & 0.008 \\ 
 50072.617 & 0.8594 & 16.052 & 0.006 &   ---  &  ---  &  0.713 & 0.007 &  0.378 & 0.007 \\ 
 50072.722 & 0.8891 & 15.895 & 0.005 &   ---  &  ---  &  0.653 & 0.009 &  0.362 & 0.009 \\ 
 50072.799 & 0.9108 & 15.792 & 0.008 &   ---  &  ---  &  0.606 & 0.010 &  0.313 & 0.011 \\ 
 50073.542 & 0.1220 & 15.709 & 0.008 &   ---  &  ---  &  0.618 & 0.011 &  0.308 & 0.011 \\ 
 50073.651 & 0.1527 & 15.769 & 0.007 &   ---  &  ---  &  0.657 & 0.009 &  0.323 & 0.010 \\ 
 50073.767 & 0.1858 & 15.817 & 0.010 &   ---  &  ---  &  0.689 & 0.011 &  0.343 & 0.012 \\ 
 50100.553 & 0.7894 & 16.241 & 0.012 &   ---  &  ---  &  0.809 & 0.034 &  0.397 & 0.018 \\ 
 50100.669 & 0.8222 & 16.183 & 0.022 &   ---  &  ---  &  0.741 & 0.036 &  0.352 & 0.035 \\ 
 50100.845 & 0.8722 & 16.015 & 0.018 &   ---  &  ---  &  0.730 & 0.024 &  0.334 & 0.022 \\ 
 50101.534 & 0.0678 & 15.678 & 0.007 &   ---  &  ---  &  0.588 & 0.021 &  0.268 & 0.016 \\ 
 50101.667 & 0.1054 & 15.728 & 0.015 &   ---  &  ---  &  0.632 & 0.020 & -0.530 & 0.018 \\ 
 50101.788 & 0.1400 & 15.776 & 0.005 &   ---  &  ---  &  0.684 & 0.008 &  0.310 & 0.009 \\ 
 50102.542 & 0.3538 & 16.008 & 0.010 &   ---  &  ---  &  0.797 & 0.038 &  0.377 & 0.013 \\ 
 50102.636 & 0.3806 & 16.026 & 0.016 &   ---  &  ---  &  0.800 & 0.038 &  0.372 & 0.025 \\ 
 50102.742 & 0.4108 & 16.045 & 0.009 &   ---  &  ---  &  0.801 & 0.023 &  0.382 & 0.017 \\ 
 50102.785 & 0.4228 & 16.054 & 0.020 &   ---  &  ---  &  0.808 & 0.034 &  0.395 & 0.029 \\ 
 50103.534 & 0.6355 & 16.173 & 0.013 &   ---  &  ---  &  0.831 & 0.021 &  0.393 & 0.020 \\ 
 50103.605 & 0.6556 & 16.176 & 0.015 &   ---  &  ---  &  0.821 & 0.033 &  0.386 & 0.025 \\ 
 50103.683 & 0.6779 & 16.198 & 0.011 &   ---  &  ---  &  0.825 & 0.021 &  0.408 & 0.019 \\ 
 50103.762 & 0.7004 & 16.190 & 0.015 &   ---  &  ---  &  0.818 & 0.024 &  0.405 & 0.026 \\ 
\enddata

\end{deluxetable}

\clearpage

\begin{deluxetable}{rrrrrrrrrr}
\tabletypesize{\scriptsize}
\tablecaption{Photometry for HV12199. \label{HV12199phot.tab}}
\tablewidth{0pt}
\tablehead{
\colhead{HJD-2400000} & \colhead{Phase}   & 
\colhead{$V$}   & \colhead{$\sigma_V$} &
\colhead{$(B-V)$}   & \colhead{$\sigma_{B-V}$} &
\colhead{$(V-I)$}   & \colhead{$\sigma_{V-I}$} &
\colhead{$(V-R)$}   & \colhead{$\sigma_{V-R}$} 
}
\startdata
 49651.841 & 0.7807 & 16.551 & 0.009 &  0.693 & 0.014 &  0.772 & 0.013 &  0.397 & 0.015 \\ 
 49668.761 & 0.1915 & 16.154 & 0.005 &  0.568 & 0.008 &  0.657 & 0.012 &  0.314 & 0.010 \\ 
 49669.609 & 0.5128 & 16.423 & 0.007 &  0.743 & 0.012 &  0.772 & 0.014 &  0.382 & 0.012 \\ 
 49669.780 & 0.5777 & 16.482 & 0.006 &  0.713 & 0.010 &  0.799 & 0.011 &  0.403 & 0.010 \\ 
 49670.795 & 0.9622 & 15.948 & 0.005 &  0.431 & 0.008 &  0.537 & 0.009 &  0.261 & 0.009 \\ 
 49671.751 & 0.3247 & 16.300 & 0.007 &  0.664 & 0.011 &  0.740 & 0.011 &  0.373 & 0.011 \\ 
 49672.769 & 0.7102 & 16.542 & 0.008 &  0.723 & 0.013 &  0.791 & 0.016 &  0.389 & 0.014 \\ 
 49713.788 & 0.2527 & 16.210 & 0.009 &  0.616 & 0.012 &  0.667 & 0.013 &  0.336 & 0.013 \\ 
 49715.760 & 0.9997 & 15.844 & 0.007 &  0.411 & 0.009 &  0.506 & 0.010 &  0.245 & 0.010 \\ 
 49716.782 & 0.3870 & 16.349 & 0.007 &  0.689 & 0.010 &  0.738 & 0.010 &  0.374 & 0.009 \\ 
 49717.658 & 0.7191 & 16.575 & 0.008 &  0.714 & 0.014 &  0.806 & 0.011 &  0.418 & 0.009 \\ 
 49717.750 & 0.7538 & 16.537 & 0.006 &  0.727 & 0.010 &  0.766 & 0.009 &  0.376 & 0.009 \\ 
 49718.701 & 0.1143 & 16.027 & 0.020 &  0.545 & 0.024 &  0.590 & 0.024 &  0.279 & 0.026 \\ 
 49718.795 & 0.1498 & 16.090 & 0.008 &  0.531 & 0.010 &  0.631 & 0.011 &  0.297 & 0.011 \\ 
 49719.628 & 0.4656 & 16.398 & 0.014 &  0.731 & 0.019 &  0.745 & 0.016 &  0.370 & 0.018 \\ 
 49719.722 & 0.5011 & 16.437 & 0.031 &  0.724 & 0.033 &  0.790 & 0.032 &  0.414 & 0.036 \\ 
 49719.799 & 0.5300 & 16.454 & 0.012 &  0.727 & 0.016 &  0.789 & 0.016 &  0.406 & 0.016 \\ 
 49720.631 & 0.8453 & 16.496 & 0.013 &  0.648 & 0.016 &  0.743 & 0.017 &  0.356 & 0.018 \\ 
 49720.713 & 0.8766 & 16.381 & 0.013 &  0.606 & 0.016 &  0.686 & 0.017 &  0.359 & 0.016 \\ 
 49720.790 & 0.9058 & 16.225 & 0.009 &  0.544 & 0.011 &  0.649 & 0.012 &  0.316 & 0.011 \\ 
 50069.590 & 0.0680 & 15.935 & 0.006 &   ---  &  ---  &  0.536 & 0.008 &  0.257 & 0.008 \\ 
 50069.757 & 0.1313 & 16.056 & 0.006 &   ---  &  ---  &  0.588 & 0.008 &  0.277 & 0.008 \\ 
 50070.575 & 0.4412 & 16.403 & 0.010 &   ---  &  ---  &  0.756 & 0.012 &  0.378 & 0.014 \\ 
 50070.690 & 0.4848 & 16.401 & 0.010 &   ---  &  ---  &  0.747 & 0.012 &  0.393 & 0.012 \\ 
 50070.831 & 0.5383 & 16.429 & 0.009 &   ---  &  ---  &  0.729 & 0.010 &  0.372 & 0.011 \\ 
 50071.565 & 0.8164 & 16.539 & 0.012 &   ---  &  ---  &  0.743 & 0.014 &  0.375 & 0.015 \\ 
 50071.694 & 0.8654 & 16.383 & 0.008 &   ---  &  ---  &  0.698 & 0.011 &  0.370 & 0.011 \\ 
 50071.795 & 0.9035 & 16.255 & 0.008 &   ---  &  ---  &  0.663 & 0.010 &  0.342 & 0.010 \\ 
 50072.617 & 0.2151 & 16.147 & 0.007 &   ---  &  ---  &  0.641 & 0.009 &   ---  &  ---  \\ 
 50072.722 & 0.2548 & 16.196 & 0.007 &   ---  &  ---  &  0.673 & 0.011 &  0.363 & 0.010 \\ 
 50072.799 & 0.2837 & 16.242 & 0.007 &   ---  &  ---  &  0.700 & 0.009 &  0.375 & 0.009 \\ 
 50073.542 & 0.5656 & 16.444 & 0.009 &   ---  &  ---  &  0.760 & 0.012 &  0.387 & 0.013 \\ 
 50073.651 & 0.6066 & 16.469 & 0.011 &   ---  &  ---  &  0.763 & 0.013 &  0.389 & 0.014 \\ 
 50073.767 & 0.6508 & 16.493 & 0.008 &   ---  &  ---  &  0.762 & 0.009 &  0.382 & 0.011 \\ 
 50100.553 & 0.8001 & 16.535 & 0.013 &   ---  &  ---  &  0.786 & 0.022 &  0.373 & 0.021 \\ 
 50100.669 & 0.8438 & 16.470 & 0.023 &   ---  &  ---  &  0.742 & 0.030 &  0.347 & 0.038 \\ 
 50100.845 & 0.9106 & 16.224 & 0.010 &   ---  &  ---  &  0.643 & 0.016 &  0.296 & 0.012 \\ 
 50101.534 & 0.1716 & 16.137 & 0.010 &   ---  &  ---  &  0.683 & 0.030 &  0.308 & 0.016 \\ 
 50101.667 & 0.2219 & 16.203 & 0.009 &   ---  &  ---  &  0.694 & 0.022 &  0.340 & 0.013 \\ 
 50101.788 & 0.2681 & 16.239 & 0.008 &   ---  &  ---  &  0.691 & 0.016 &  0.329 & 0.011 \\ 
 50102.542 & 0.5535 & 16.455 & 0.012 &   ---  &  ---  &  0.778 & 0.029 &  0.377 & 0.019 \\ 
 50102.636 & 0.5892 & 16.471 & 0.018 &   ---  &  ---  &  0.780 & 0.029 &  0.380 & 0.028 \\ 
 50102.742 & 0.6295 & 16.498 & 0.012 &   ---  &  ---  &  0.796 & 0.018 &  0.386 & 0.018 \\ 
 50102.785 & 0.6455 & 16.491 & 0.014 &   ---  &  ---  &  0.770 & 0.021 &  0.378 & 0.022 \\ 
 50103.534 & 0.9295 & 16.091 & 0.012 &   ---  &  ---  &  0.615 & 0.017 &  0.284 & 0.018 \\ 
 50103.605 & 0.9563 & 15.948 & 0.015 &   ---  &  ---  &  0.582 & 0.026 &  0.270 & 0.022 \\ 
 50103.683 & 0.9860 & 15.866 & 0.011 &   ---  &  ---  &  0.521 & 0.014 &  0.244 & 0.015 \\ 
 50103.762 & 0.0160 & 15.866 & 0.011 &   ---  &  ---  &  0.489 & 0.017 &  0.237 & 0.016 \\ 
\enddata

\end{deluxetable}

\clearpage

\begin{deluxetable}{rrrrrrrrrr}
\tabletypesize{\scriptsize}
\tablecaption{Photometry for HV12202. \label{HV12202phot.tab}}
\tablewidth{0pt}
\tablehead{
\colhead{HJD-2400000} & \colhead{Phase}   & 
\colhead{$V$}   & \colhead{$\sigma_V$} &
\colhead{$(B-V)$}   & \colhead{$\sigma_{B-V}$} &
\colhead{$(V-I)$}   & \colhead{$\sigma_{V-I}$} &
\colhead{$(V-R)$}   & \colhead{$\sigma_{V-R}$} 
}
\startdata
 49651.841 & 0.6558 & 16.281 & 0.009 &  0.735 & 0.013 &  0.762 & 0.015 &  0.387 & 0.015 \\ 
 49668.761 & 0.1117 & 15.834 & 0.010 &  0.578 & 0.015 &  0.591 & 0.016 &  0.359 & 0.016 \\ 
 49669.609 & 0.3851 & 16.148 & 0.010 &  0.696 & 0.016 &  0.758 & 0.015 &  0.381 & 0.016 \\ 
 49669.780 & 0.4404 & 16.191 & 0.010 &  0.713 & 0.014 &  0.756 & 0.015 &  0.375 & 0.016 \\ 
 49670.795 & 0.7675 & 16.315 & 0.010 &  0.690 & 0.014 &  0.767 & 0.016 &  0.411 & 0.016 \\ 
 49671.751 & 0.0760 & 15.845 & 0.007 &  0.535 & 0.011 &  0.610 & 0.011 &  0.304 & 0.012 \\ 
 49672.769 & 0.4041 & 16.179 & 0.007 &  0.689 & 0.011 &  0.777 & 0.013 &  0.396 & 0.013 \\ 
 49713.788 & 0.6314 & 16.279 & 0.007 &  0.721 & 0.011 &  0.763 & 0.012 &  0.556 & 0.011 \\ 
 49714.750 & 0.9416 & 15.823 & 0.016 &  0.688 & 0.022 &   ---  &  ---  &  0.160 & 0.027 \\ 
 49715.760 & 0.2672 & 16.044 & 0.007 &  0.657 & 0.009 &  0.701 & 0.011 &  0.333 & 0.010 \\ 
 49717.658 & 0.8794 & 15.995 & 0.007 &  0.655 & 0.010 &   ---  &  ---  &  0.226 & 0.011 \\ 
 49717.750 & 0.9090 & 16.010 & 0.005 &  0.560 & 0.007 &  0.632 & 0.010 &  0.296 & 0.012 \\ 
 49718.795 & 0.2460 & 16.044 & 0.006 &  0.656 & 0.008 &  0.700 & 0.011 &  0.334 & 0.009 \\ 
 49719.722 & 0.5449 & 16.178 & 0.030 &  0.778 & 0.031 &   ---  &  ---  &  0.338 & 0.035 \\ 
 49719.799 & 0.5696 & 16.254 & 0.011 &  0.708 & 0.014 &  0.753 & 0.016 &   ---  &  ---  \\ 
 49720.631 & 0.8378 & 16.231 & 0.010 &  0.647 & 0.012 &  0.721 & 0.013 &  0.351 & 0.013 \\ 
 49720.713 & 0.8645 & 16.170 & 0.010 &  0.625 & 0.013 &  0.691 & 0.013 &  0.369 & 0.013 \\ 
 49720.790 & 0.8893 & 16.066 & 0.007 &  0.591 & 0.009 &  0.637 & 0.010 &  0.325 & 0.009 \\ 
 50069.590 & 0.3648 & 16.130 & 0.015 &   ---  &  ---  &   ---  &  ---  &  0.376 & 0.019 \\ 
 50069.757 & 0.4186 & 16.176 & 0.017 &   ---  &  ---  &  0.763 & 0.019 &  0.355 & 0.024 \\ 
 50070.575 & 0.6824 & 16.284 & 0.017 &   ---  &  ---  &   ---  &  ---  &   ---  &  ---  \\ 
 50070.690 & 0.7195 & 16.249 & 0.012 &   ---  &  ---  &  0.677 & 0.013 &  0.225 & 0.020 \\ 
 50070.831 & 0.7650 & 16.285 & 0.014 &   ---  &  ---  &   ---  &  ---  &   ---  &  ---  \\ 
 50071.565 & 0.0017 & 15.750 & 0.010 &   ---  &  ---  &  0.520 & 0.013 &  0.222 & 0.014 \\ 
 50071.795 & 0.0758 & 15.792 & 0.007 &   ---  &  ---  &  0.548 & 0.009 &  0.258 & 0.010 \\ 
 50072.617 & 0.3410 & 16.085 & 0.009 &   ---  &  ---  &  0.716 & 0.013 &   ---  &  ---  \\ 
 50072.722 & 0.3748 & 16.115 & 0.016 &   ---  &  ---  &   ---  &  ---  &   ---  &  ---  \\ 
 50072.799 & 0.3994 & 16.115 & 0.013 &   ---  &  ---  &   ---  &  ---  &  0.386 & 0.020 \\ 
 50073.542 & 0.6393 & 16.240 & 0.009 &   ---  &  ---  &  0.727 & 0.012 &  0.354 & 0.013 \\ 
 50073.651 & 0.6742 & 16.279 & 0.009 &   ---  &  ---  &  0.749 & 0.011 &  0.388 & 0.011 \\ 
 50100.553 & 0.3493 & 16.067 & 0.043 &   ---  &  ---  &  0.804 & 0.098 &  0.379 & 0.067 \\ 
 50100.669 & 0.3865 & 16.075 & 0.073 &   ---  &  ---  &  0.679 & 0.083 &  0.346 & 0.117 \\ 
 50100.845 & 0.4433 & 16.096 & 0.016 &   ---  &  ---  &  0.870 & 0.024 &  0.395 & 0.019 \\ 
 50101.534 & 0.6655 & 16.221 & 0.041 &   ---  &  ---  &  0.752 & 0.070 &  0.298 & 0.047 \\ 
 50101.667 & 0.7083 & 16.246 & 0.034 &   ---  &  ---  &  0.719 & 0.038 &  0.426 & 0.057 \\ 
 50101.788 & 0.7476 & 16.225 & 0.027 &   ---  &  ---  &  0.819 & 0.036 &  0.381 & 0.044 \\ 
 50102.542 & 0.9904 & 15.771 & 0.038 &   ---  &  ---  &  0.685 & 0.080 &  0.282 & 0.060 \\ 
 50102.636 & 0.0209 & 15.767 & 0.037 &   ---  &  ---  &  0.665 & 0.092 &  0.304 & 0.079 \\ 
 50102.742 & 0.0552 & 15.776 & 0.023 &   ---  &  ---  &  0.647 & 0.048 &  0.312 & 0.031 \\ 
 50102.785 & 0.0688 & 15.775 & 0.032 &   ---  &  ---  &  0.660 & 0.049 &  0.307 & 0.039 \\ 
 50103.534 & 0.3104 & 16.026 & 0.043 &   ---  &  ---  &  0.696 & 0.050 &  0.380 & 0.067 \\ 
 50103.605 & 0.3333 & 16.024 & 0.056 &   ---  &  ---  &  0.697 & 0.069 &  0.283 & 0.065 \\ 
 50103.683 & 0.3586 & 16.046 & 0.037 &   ---  &  ---  &  0.767 & 0.067 &  0.366 & 0.050 \\ 
 50103.762 & 0.3841 & 16.037 & 0.027 &   ---  &  ---  &  0.800 & 0.039 &  0.379 & 0.035 \\ 
\enddata

\end{deluxetable}

\clearpage

\begin{deluxetable}{rrrrrrrrrr}
\tabletypesize{\scriptsize}
\tablecaption{Photometry for HV12203. \label{HV12203phot.tab}}
\tablewidth{0pt}
\tablehead{
\colhead{HJD-2400000} & \colhead{Phase}   & 
\colhead{$V$}   & \colhead{$\sigma_V$} &
\colhead{$(B-V)$}   & \colhead{$\sigma_{B-V}$} &
\colhead{$(V-I)$}   & \colhead{$\sigma_{V-I}$} &
\colhead{$(V-R)$}   & \colhead{$\sigma_{V-R}$} 
}
\startdata
 49651.841 & 0.8680 & 16.196 & 0.009 &  0.626 & 0.014 &  0.686 & 0.015 &  0.331 & 0.015 \\ 
 49668.761 & 0.5954 & 16.363 & 0.006 &  0.756 & 0.010 &  0.808 & 0.013 &  0.412 & 0.011 \\ 
 49669.609 & 0.8824 & 16.112 & 0.008 &  0.596 & 0.013 &  0.646 & 0.014 &  0.312 & 0.014 \\ 
 49669.780 & 0.9404 & 15.899 & 0.011 &  0.484 & 0.016 &  0.563 & 0.016 &  0.273 & 0.016 \\ 
 49670.795 & 0.2839 & 16.150 & 0.008 &  0.637 & 0.012 &  0.748 & 0.014 &  0.414 & 0.014 \\ 
 49671.751 & 0.6077 & 16.349 & 0.007 &  0.744 & 0.011 &  0.787 & 0.012 &  0.411 & 0.012 \\ 
 49672.769 & 0.9522 & 15.862 & 0.006 &  0.467 & 0.009 &  0.556 & 0.013 &  0.278 & 0.011 \\ 
 49713.788 & 0.8377 & 16.295 & 0.007 &  0.657 & 0.012 &  0.734 & 0.011 &  0.367 & 0.011 \\ 
 49715.760 & 0.5051 & 16.265 & 0.006 &  0.744 & 0.009 &  0.752 & 0.008 &  0.357 & 0.008 \\ 
 49716.782 & 0.8511 & 16.255 & 0.005 &  0.672 & 0.009 &  0.699 & 0.008 &  0.352 & 0.006 \\ 
 49717.658 & 0.1478 & 15.992 & 0.005 &  0.559 & 0.007 &  0.655 & 0.007 &  0.325 & 0.006 \\ 
 49717.750 & 0.1788 & 15.999 & 0.004 &  0.606 & 0.006 &  0.634 & 0.006 &  0.309 & 0.006 \\ 
 49718.701 & 0.5009 & 16.289 & 0.016 &  0.751 & 0.019 &  0.766 & 0.019 &  0.366 & 0.019 \\ 
 49718.795 & 0.5326 & 16.293 & 0.006 &  0.738 & 0.009 &  0.746 & 0.009 &  0.369 & 0.008 \\ 
 49719.628 & 0.8147 & 16.334 & 0.012 &  0.706 & 0.017 &  0.727 & 0.014 &  0.358 & 0.015 \\ 
 49719.722 & 0.8464 & 16.249 & 0.028 &  0.680 & 0.030 &  0.702 & 0.029 &  0.357 & 0.034 \\ 
 49719.799 & 0.8723 & 16.204 & 0.010 &  0.631 & 0.013 &  0.689 & 0.013 &  0.357 & 0.013 \\ 
 49720.631 & 0.1539 & 15.964 & 0.009 &  0.590 & 0.011 &  0.635 & 0.012 &  0.303 & 0.011 \\ 
 49720.713 & 0.1819 & 16.012 & 0.010 &  0.598 & 0.013 &  0.624 & 0.013 &  0.324 & 0.012 \\ 
 49720.790 & 0.2080 & 16.026 & 0.007 &  0.607 & 0.010 &  0.658 & 0.010 &  0.324 & 0.009 \\ 
 50069.590 & 0.2807 & 16.112 & 0.007 &   ---  &  ---  &  0.699 & 0.009 &  0.385 & 0.009 \\ 
 50069.757 & 0.3372 & 16.181 & 0.008 &   ---  &  ---  &  0.733 & 0.010 &  0.410 & 0.011 \\ 
 50070.575 & 0.6141 & 16.340 & 0.008 &   ---  &  ---  &  0.783 & 0.009 &  0.414 & 0.011 \\ 
 50070.690 & 0.6531 & 16.332 & 0.008 &   ---  &  ---  &  0.754 & 0.009 &  0.381 & 0.009 \\ 
 50070.831 & 0.7009 & 16.370 & 0.008 &   ---  &  ---  &  0.750 & 0.009 &  0.392 & 0.009 \\ 
 50071.565 & 0.9494 & 15.872 & 0.006 &   ---  &  ---  &  0.575 & 0.008 &  0.273 & 0.008 \\ 
 50071.694 & 0.9931 & 15.759 & 0.007 &   ---  &  ---  &  0.520 & 0.011 &  0.274 & 0.010 \\ 
 50071.795 & 0.0271 & 15.809 & 0.008 &   ---  &  ---  &  0.551 & 0.010 &  0.275 & 0.009 \\ 
 50072.617 & 0.3055 & 16.120 & 0.007 &   ---  &  ---  &  0.705 & 0.009 &  0.391 & 0.009 \\ 
 50072.722 & 0.3410 & 16.177 & 0.006 &   ---  &  ---  &  0.756 & 0.011 &  0.424 & 0.009 \\ 
 50072.799 & 0.3668 & 16.185 & 0.007 &   ---  &  ---  &  0.738 & 0.009 &  0.400 & 0.009 \\ 
 50073.542 & 0.6186 & 16.337 & 0.008 &   ---  &  ---  &  0.769 & 0.011 &  0.407 & 0.011 \\ 
 50073.651 & 0.6553 & 16.346 & 0.010 &   ---  &  ---  &  0.766 & 0.012 &  0.384 & 0.013 \\ 
 50073.767 & 0.6948 & 16.372 & 0.008 &   ---  &  ---  &  0.770 & 0.009 &  0.385 & 0.009 \\ 
 50100.553 & 0.7621 & 16.390 & 0.014 &   ---  &  ---  &  0.785 & 0.042 &  0.374 & 0.026 \\ 
 50100.669 & 0.8011 & 16.364 & 0.023 &   ---  &  ---  &  0.738 & 0.045 &  0.357 & 0.043 \\ 
 50100.845 & 0.8608 & 16.237 & 0.014 &   ---  &  ---  &  0.772 & 0.025 &  0.347 & 0.018 \\ 
 50101.534 & 0.0940 & 15.903 & 0.008 &   ---  &  ---  &  0.567 & 0.024 &  0.280 & 0.021 \\ 
 50101.667 & 0.1389 & 15.975 & 0.008 &   ---  &  ---  &  0.648 & 0.017 &  0.318 & 0.014 \\ 
 50101.788 & 0.1802 & 16.012 & 0.008 &   ---  &  ---  &  0.693 & 0.009 &  0.312 & 0.014 \\ 
 50102.542 & 0.4351 & 16.247 & 0.011 &   ---  &  ---  &  0.760 & 0.047 &  0.364 & 0.016 \\ 
 50102.636 & 0.4671 & 16.280 & 0.018 &   ---  &  ---  &  0.785 & 0.054 &  0.382 & 0.037 \\ 
 50102.742 & 0.5031 & 16.283 & 0.011 &   ---  &  ---  &  0.830 & 0.033 &  0.385 & 0.017 \\ 
 50102.785 & 0.5174 & 16.287 & 0.021 &   ---  &  ---  &  0.796 & 0.037 &  0.376 & 0.029 \\ 
 50103.534 & 0.7710 & 16.377 & 0.014 &   ---  &  ---  &  0.795 & 0.028 &  0.375 & 0.024 \\ 
 50103.605 & 0.7950 & 16.353 & 0.019 &   ---  &  ---  &  0.773 & 0.047 &  0.384 & 0.030 \\ 
 50103.683 & 0.8216 & 16.320 & 0.017 &   ---  &  ---  &  0.777 & 0.030 &  0.364 & 0.025 \\ 
 50103.762 & 0.8484 & 16.230 & 0.017 &   ---  &  ---  &  0.745 & 0.028 &  0.355 & 0.027 \\ 
\enddata

\end{deluxetable}

\clearpage

\begin{deluxetable}{rrrrrrrrrr}
\tabletypesize{\scriptsize}
\tablecaption{Photometry for HV12204. \label{HV12204phot.tab}}
\tablewidth{0pt}
\tablehead{
\colhead{HJD-2400000} & \colhead{Phase}   & 
\colhead{$V$}   & \colhead{$\sigma_V$} &
\colhead{$(B-V)$}   & \colhead{$\sigma_{B-V}$} &
\colhead{$(V-I)$}   & \colhead{$\sigma_{V-I}$} &
\colhead{$(V-R)$}   & \colhead{$\sigma_{V-R}$} 
}
\startdata
 49651.841 & 0.5467 & 15.957 & 0.006 &  0.645 & 0.009 &  0.751 & 0.010 &  0.378 & 0.010 \\ 
 49668.761 & 0.4669 & 15.898 & 0.007 & -0.354 & 0.011 &  0.730 & 0.011 &  0.384 & 0.011 \\ 
 49669.609 & 0.7135 & 16.056 & 0.006 &  0.658 & 0.010 &  0.760 & 0.012 &  0.378 & 0.010 \\ 
 49669.780 & 0.7633 & 16.083 & 0.010 &  0.664 & 0.014 &  0.734 & 0.013 &  0.352 & 0.016 \\ 
 49670.795 & 0.0584 & 15.303 & 0.005 &  0.333 & 0.007 &  0.456 & 0.009 &  0.243 & 0.009 \\ 
 49671.751 & 0.3366 & 15.733 & 0.004 &  0.574 & 0.007 &  0.686 & 0.009 &  0.344 & 0.007 \\ 
 49672.769 & 0.6325 & 16.033 & 0.006 &  0.639 & 0.010 &  0.773 & 0.010 &  0.386 & 0.010 \\ 
 49713.788 & 0.5610 & 15.932 & 0.007 &  0.658 & 0.011 &  0.725 & 0.012 &  0.353 & 0.011 \\ 
 49715.760 & 0.1343 & 15.418 & 0.006 &  0.438 & 0.008 &  0.516 & 0.008 & -0.161 & 0.008 \\ 
 49716.782 & 0.4316 & 15.840 & 0.005 &  0.654 & 0.007 &  0.692 & 0.008 &  0.346 & 0.007 \\ 
 49717.658 & 0.6865 & 16.028 & 0.006 &  0.695 & 0.008 &  0.726 & 0.008 &  0.368 & 0.008 \\ 
 49717.750 & 0.7131 & 16.044 & 0.005 &  0.679 & 0.007 &  0.744 & 0.008 &  0.366 & 0.008 \\ 
 49718.701 & 0.9898 & 15.176 & 0.017 &  0.342 & 0.020 &  0.351 & 0.020 &  0.139 & 0.021 \\ 
 49718.795 & 0.0170 & 15.170 & 0.005 &  0.331 & 0.007 &  0.387 & 0.008 &  0.158 & 0.008 \\ 
 49719.628 & 0.2594 & 15.634 & 0.012 &  0.600 & 0.018 &  0.620 & 0.014 &  0.292 & 0.015 \\ 
 49719.722 & 0.2866 & 15.620 & 0.019 &  0.582 & 0.020 &  0.607 & 0.020 &  0.267 & 0.023 \\ 
 49719.799 & 0.3089 & 15.662 & 0.008 &  0.603 & 0.010 &  0.620 & 0.011 &  0.288 & 0.011 \\ 
 49720.631 & 0.5508 & 15.931 & 0.010 &  0.679 & 0.012 &  0.712 & 0.013 &  0.344 & 0.013 \\ 
 49720.713 & 0.5748 & 15.955 & 0.010 &  0.663 & 0.012 &  0.733 & 0.013 &  0.365 & 0.012 \\ 
 49720.790 & 0.5972 & 15.959 & 0.006 &  0.663 & 0.008 &  0.729 & 0.009 &  0.358 & 0.008 \\ 
 50069.590 & 0.0291 & 15.177 & 0.007 &   ---  &  ---  &  0.378 & 0.009 &  0.179 & 0.009 \\ 
 50069.757 & 0.0777 & 15.324 & 0.010 &   ---  &  ---  &  0.457 & 0.012 &  0.221 & 0.013 \\ 
 50070.575 & 0.3155 & 15.709 & 0.008 &   ---  &  ---  &  0.659 & 0.009 &  0.319 & 0.011 \\ 
 50070.690 & 0.3490 & 15.689 & 0.008 &   ---  &  ---  &  0.591 & 0.009 &  0.306 & 0.010 \\ 
 50070.831 & 0.3901 & 15.770 & 0.009 &   ---  &  ---  &  0.644 & 0.010 &  0.334 & 0.011 \\ 
 50071.565 & 0.6035 & 15.989 & 0.006 &   ---  &  ---  &  0.739 & 0.008 &  0.376 & 0.010 \\ 
 50071.694 & 0.6411 & 15.972 & 0.009 &   ---  &  ---  &  0.703 & 0.013 &  0.360 & 0.011 \\ 
 50071.795 & 0.6704 & 16.041 & 0.007 &   ---  &  ---  &  0.748 & 0.009 &  0.384 & 0.009 \\ 
 50072.617 & 0.9095 & 15.749 & 0.005 &   ---  &  ---  &  0.598 & 0.007 &  0.319 & 0.008 \\ 
 50072.722 & 0.9400 & 15.454 & 0.008 &   ---  &  ---  &  0.465 & 0.012 &  0.267 & 0.011 \\ 
 50072.799 & 0.9622 & 15.255 & 0.008 &   ---  &  ---  &  0.395 & 0.009 &  0.216 & 0.011 \\ 
 50073.542 & 0.1785 & 15.494 & 0.009 &   ---  &  ---  &  0.543 & 0.011 &  0.278 & 0.011 \\ 
 50073.651 & 0.2100 & 15.557 & 0.007 &   ---  &  ---  &  0.592 & 0.009 &  0.293 & 0.010 \\ 
 50073.767 & 0.2439 & 15.611 & 0.011 &   ---  &  ---  &  0.606 & 0.012 &  0.311 & 0.014 \\ 
 50100.553 & 0.0333 & 15.224 & 0.014 &   ---  &  ---  &  0.456 & 0.044 &  0.202 & 0.025 \\ 
 50100.669 & 0.0668 & 15.307 & 0.024 &   ---  &  ---  &  0.451 & 0.054 &  0.183 & 0.047 \\ 
 50100.845 & 0.1181 & 15.415 & 0.015 &   ---  &  ---  &  0.587 & 0.027 &  0.243 & 0.019 \\ 
 50101.534 & 0.3184 & 15.716 & 0.008 &   ---  &  ---  &  0.623 & 0.030 &  0.319 & 0.020 \\ 
 50101.667 & 0.3570 & 15.766 & 0.012 &   ---  &  ---  &  0.694 & 0.030 &  0.350 & 0.016 \\ 
 50101.788 & 0.3924 & 15.807 & 0.008 &   ---  &  ---  &  0.725 & 0.011 &  0.334 & 0.014 \\ 
 50102.542 & 0.6115 & 15.979 & 0.014 &   ---  &  ---  &  0.788 & 0.050 &  0.359 & 0.017 \\ 
 50102.636 & 0.6389 & 16.006 & 0.019 &   ---  &  ---  &  0.764 & 0.055 &  0.372 & 0.038 \\ 
 50102.742 & 0.6698 & 16.019 & 0.012 &   ---  &  ---  &  0.762 & 0.038 &  0.374 & 0.018 \\ 
 50102.785 & 0.6821 & 16.029 & 0.022 &   ---  &  ---  &  0.765 & 0.040 &  0.372 & 0.031 \\ 
 50103.534 & 0.9000 & 15.847 & 0.015 &   ---  &  ---  &  0.683 & 0.033 &  0.307 & 0.025 \\ 
 50103.683 & 0.9434 & 15.452 & 0.015 &   ---  &  ---  &  0.512 & 0.037 &  0.242 & 0.021 \\ 
 50103.762 & 0.9665 & 15.235 & 0.020 &   ---  &  ---  &  0.461 & 0.034 &  0.204 & 0.029 \\ 
\enddata

\end{deluxetable}

\clearpage

\begin{deluxetable}{rrrrrrrrrr}
\tabletypesize{\scriptsize}
\tablecaption{Photometry for NGC1866-V7. \label{V7phot.tab}}
\tablewidth{0pt}
\tablehead{
\colhead{HJD-2400000} & \colhead{Phase}   & 
\colhead{$V$}   & \colhead{$\sigma_V$} &
\colhead{$(B-V)$}   & \colhead{$\sigma_{B-V}$} &
\colhead{$(V-I)$}   & \colhead{$\sigma_{V-I}$} &
\colhead{$(V-R)$}   & \colhead{$\sigma_{V-R}$} 
}
\startdata
 49713.788 & 0.4102 & 16.027 & 0.009 &  0.716 & 0.013 &  0.736 & 0.013 &  0.357 & 0.013 \\ 
 49714.750 & 0.6941 & 16.293 & 0.025 &  0.582 & 0.030 &  0.821 & 0.030 &  0.477 & 0.031 \\ 
 49715.760 & 0.9921 & 15.792 & 0.007 &  0.543 & 0.011 &  0.621 & 0.010 &  0.314 & 0.010 \\ 
 49716.782 & 0.2938 & 15.968 & 0.005 &  0.628 & 0.010 &   ---  &  ---  &  0.370 & 0.007 \\ 
 49717.658 & 0.5524 & 16.074 & 0.007 &  0.693 & 0.010 &  0.731 & 0.009 &   ---  &  ---  \\ 
 49717.750 & 0.5795 & 16.086 & 0.006 &  0.690 & 0.010 &  0.733 & 0.009 &  0.362 & 0.008 \\ 
 49718.701 & 0.8602 & 16.119 & 0.017 &  0.553 & 0.021 &  0.756 & 0.020 &  0.400 & 0.021 \\ 
 49718.795 & 0.8879 & 15.988 & 0.006 &  0.644 & 0.009 &  0.691 & 0.009 &  0.328 & 0.008 \\ 
 49719.628 & 0.1339 & 15.921 & 0.013 &  0.528 & 0.018 &  0.701 & 0.015 &  0.366 & 0.015 \\ 
 49719.722 & 0.1615 & 15.974 & 0.025 &  0.461 & 0.026 &  0.784 & 0.026 &  0.409 & 0.030 \\ 
 49719.799 & 0.1841 & 15.920 & 0.009 &  0.523 & 0.011 &  0.719 & 0.012 &  0.364 & 0.012 \\ 
 49720.631 & 0.4296 & 16.043 & 0.009 &  0.659 & 0.013 &  0.748 & 0.012 &  0.370 & 0.012 \\ 
 49720.713 & 0.4540 & 16.054 & 0.010 &  0.664 & 0.014 &  0.761 & 0.013 &  0.389 & 0.013 \\ 
 49720.790 & 0.4767 & 16.043 & 0.007 &  0.684 & 0.013 &  0.748 & 0.011 &  0.368 & 0.009 \\ 
 50069.590 & 0.4170 & 16.021 & 0.008 &   ---  &  ---  &   ---  &  ---  &  0.361 & 0.011 \\ 
 50069.757 & 0.4663 & 16.067 & 0.008 &   ---  &  ---  &   ---  &  ---  &  0.362 & 0.010 \\ 
 50070.575 & 0.7077 & 16.078 & 0.009 &   ---  &  ---  &   ---  &  ---  &  0.357 & 0.011 \\ 
 50070.690 & 0.7417 & 16.030 & 0.008 &   ---  &  ---  &   ---  &  ---  &  0.353 & 0.013 \\ 
 50070.831 & 0.7834 & 15.996 & 0.010 &   ---  &  ---  &   ---  &  ---  &  0.369 & 0.011 \\ 
 50071.565 & 0.0000 & 15.792 & 0.006 &   ---  &  ---  &   ---  &  ---  &  0.301 & 0.010 \\ 
 50071.694 & 0.0381 & 15.797 & 0.004 &   ---  &  ---  &   ---  &  ---  &   ---  &  ---  \\ 
 50071.795 & 0.0678 & 15.854 & 0.009 &   ---  &  ---  &  0.660 & 0.010 &  0.338 & 0.011 \\ 
 50072.617 & 0.3105 & 15.942 & 0.003 &   ---  &  ---  &  0.748 & 0.031 &  0.339 & 0.008 \\ 
 50072.722 & 0.3414 & 15.995 & 0.007 &   ---  &  ---  &   ---  &  ---  &  0.402 & 0.010 \\ 
 50072.799 & 0.3640 & 15.996 & 0.007 &   ---  &  ---  &  0.708 & 0.010 &  0.365 & 0.009 \\ 
 50073.542 & 0.5835 & 16.051 & 0.007 &   ---  &  ---  &  0.702 & 0.011 &  0.335 & 0.011 \\ 
 50073.651 & 0.6155 & 16.099 & 0.011 &   ---  &  ---  &  0.732 & 0.013 &  0.371 & 0.013 \\ 
 50073.767 & 0.6499 & 16.100 & 0.007 &   ---  &  ---  &   ---  &  ---  &  0.347 & 0.009 \\ 
\enddata

\end{deluxetable}

\begin{deluxetable}{rllll}
\tablecaption{The new ephemerides adopted for the stars. \label{ephem.tab}}
\tablewidth{0pt}
\tablehead{
\colhead{} & \colhead{Period} & \colhead{$\sigma(\mbox{Period})$} & 
\colhead{Epoch} & \colhead{$\sigma(\mbox{Epoch})$} \\
\colhead{Identifier} & \colhead{(days)} & \colhead{(days)}  & 
\colhead{(HJD)} & \colhead{(days)}
}
\startdata
HV~12197    & 3.14381 & 0.00002 & 2450071.72  & 0.01 \\
HV~12198    & 3.52279 & 0.00001 & 2450069.59  & 0.01 \\
HV~12199    & 2.63918 & 0.00001 & 2450103.72  & 0.01 \\
HV~12202    & 3.10112 & 0.00003 & 2450071.56  & 0.01 \\
HV~12203    & 2.95411 & 0.00001 & 2450071.715 & 0.005 \\
HV~12204    & 3.43876 & 0.00001 & 2450069.49  & 0.01 \\
NGC~1866-V7 & 3.38837 & 0.045   & 2450071.565 & 0.005 \\
\enddata

\end{deluxetable}

\end{document}